\documentclass[conference]{IEEEtran}
\IEEEoverridecommandlockouts
\usepackage {booktabs}
\usepackage{url}
\usepackage{hyperref}

\usepackage{cite}
\usepackage{amsmath,amssymb,amsfonts}
\usepackage{algorithmic}
\usepackage{graphicx}
\usepackage{textcomp}
\usepackage{xcolor}
\def\BibTeX{{\rm B\kern-.05em{\sc i\kern-.025em b}\kern-.08em
    T\kern-.1667em\lower.7ex\hbox{E}\kern-.125emX}}
\begin{document}

\title{Exploring Pre-trained General-purpose Audio Representations for Heart Murmur Detection}

%\author{\IEEEauthorblockN{Daisuke Niizumi, Daiki Takeuchi, Yasunori Ohishi, Noboru Harada, and Kunio Kashino}
%\IEEEauthorblockA{\textit{NTT Communication Science Laboratories} \\
%\textit{NTT Corporation}\\
%Atsugi, Japan \\
%E-mail: daisuke.niizumi@ntt.com}
%}
\author{Daisuke~Niizumi, Daiki~Takeuchi, Yasunori~Ohishi,~\IEEEmembership{Member,~IEEE,} \\
Noboru~Harada,~\IEEEmembership{Senior~Member,~IEEE,} and~Kunio~Kashino,~\IEEEmembership{Senior~Member,~IEEE}
        % <-this % stops a space
\thanks{The authors are with Communication Science  Laboratories, Nippon Telegraph and Telephone Corporation,  Atsugi 243-0198, Japan (e-mail: daisuke.niizumi@ntt.com; d.takeuchi@ntt.com; yasunori.ohishi@ntt.com; harada.noboru@ntt.com; kunio.kashino@ntt.com)}
}

\maketitle

\begin{abstract}
To reduce the need for skilled clinicians in heart sound interpretation, recent studies on automating cardiac auscultation have explored deep learning approaches.
However, despite the demands for large data for deep learning, the size of the heart sound datasets is limited, and no pre-trained model is available.
On the contrary, many pre-trained models for general audio tasks are available as general-purpose audio representations.
This study explores the potential of general-purpose audio representations pre-trained on large-scale datasets for transfer learning in heart murmur detection.
Experiments on the CirCor DigiScope heart sound dataset show that the recent self-supervised learning Masked Modeling Duo (M2D) outperforms previous methods with the results of a weighted accuracy of 0.832 and an unweighted average recall of 0.713. Experiments further confirm improved performance by ensembling M2D with other models.
These results demonstrate the effectiveness of general-purpose audio representation in processing heart sounds and open the way for further applications\footnote{Our code is available online which runs on a 24 GB consumer GPU: \url{https://github.com/nttcslab/m2d/tree/master/app/circor}}.
\end{abstract}

\begin{IEEEkeywords}
cardiac auscultation, heart murmur detection, general-purpose audio representation, transfer learning
\end{IEEEkeywords}

\section{Introduction}
Heart sound auscultation is an accessible diagnostic screening tool for identifying heart murmurs. However, interpreting heart sounds requires a skilled clinician, and automated algorithmic approaches are thus being explored to reduce the burden of training professionals\cite{survey_ren2023comprehensive,survey_Zhao2023review}.
In the 2022 George B. Moody PhysioNet Challenge\cite{GeorgeBMoodyPhysioNetChallenge2022,PhysioNet}, which encouraged developments for detecting heart murmurs and abnormal cardiac function, many proposals took deep learning approaches.

Deep learning models generally require a large amount of data for training; however, the size of the heart sound dataset is limited. Although the CirCor DigiScope dataset\cite{CirCor2022} used in the challenge is the largest heart sound dataset, it contains only 3,163 publicly available recordings and is thus considered undersized for applying deep learning, especially for modern transformer models.

\begin{table}[tb!]
%\vspace{-10pt}
\caption{Explored general-purpose audio representations\\ and their training settings}
\label{tab:exp:models-settings}
%\vspace{-5pt}
\centering
\resizebox{1.0\columnwidth}{!}{%
\begin{tabular}{lcccc}
\toprule
& CNN14\cite{kong2020panns} & BYOL-A\cite{niizumi2021byol-a} & AST\cite{gong2021ast} & M2D\cite{niizumi2022M2D} \\
\midrule
\multicolumn{5}{l}{\textit{(i) Model/pre-training configurations}}  \\
Architecture & \multicolumn{2}{c}{CNN} & \multicolumn{2}{c}{Transformer} \\
Pre-training$^\dagger$ & SL & SSL & SL & SSL \\
Parameters & 79.9M & 5.3M & 87.3M & 85.4M \\
Feature dimensions & 2,048-d & 2,048-d & 768-d & 3,840-d \\
\midrule
\multicolumn{5}{l}{\textit{(ii) Training settings for fine-tuning on the CirCor DigiScope dataset}}  \\
Learning rate & 0.001 & 0.001 & 0.00003 & 0.00025 \\
Batch size & 256 & 256 & 64 & 32 \\
SpecAugment$^\ddagger$\cite{specaugment} & 20/200 & 20/50 & 40/100 & 0/0 \\
\bottomrule
\addlinespace[0.05cm]
\multicolumn{5}{l}{$^{\dagger}$ \scriptsize{Supervised learning (SL) or self-supervised learning (SSL).}} \\
\multicolumn{5}{l}{$^{\ddagger}$ \scriptsize{Parameters for frequency and time masking each.}} \\
\end{tabular}
}
\vspace{-5pt}
\end{table}

Transfer learning is the common approach for applying deep learning to such a small dataset.
In transfer learning, a large-scale dataset close to the application domain is used to pre-train models to obtain effective representations, and then the pre-trained models are transferred to the application task.
Previous studies have evaluated existing pre-trained models, including those from the image domain\cite{Koike2020CNN14,Pathak2022Ensembled}. However, recent studies, such as those represented at the abovementioned challenge, have rarely used transfer learning.

On the other hand, many pre-trained models have recently been proposed as general-purpose audio representations\cite{saeed2020cola,niizumi2021byol-a,niizumi2022M2D}. They are self-supervised learning models typically pre-trained on AudioSet\cite{gemmeke2017audioset}, a large-scale dataset of YouTube video sounds, and have shown effectiveness on diverse tasks, including environmental sounds, speech, and music. In addition, various supervised learning models pre-trained on AudioSet have also shown effectiveness on various tasks\cite{kong2020panns,gong2021ast}, and they can also be considered implicit general-purpose audio representations. However, the effectiveness of these representations with heart sounds has yet to be evaluated.

This study explores the effectiveness of general-purpose audio representations for heart sounds. We use the heart murmur detection task from the 2022 PhysioNet Challenge because, while previous studies have not compared results with state-of-the-art (SOTA) performance such as \cite{Koike2020CNN14}, SOTA results are available with this task\cite{Panah2023Exploring}, which allows us to assess the top performances. For the general-purpose audio representations, we evaluate SOTA pre-trained models listed in Table \ref{tab:exp:models-settings}, chosen for the diversity of learning methods and model architectures.

Experimental results show that the latest general-purpose audio representation, Masked Modeling Duo (M2D)\cite{niizumi2022M2D}, outperformed the SOTA results. Other models also showed different trends in class detection performance from M2D, and ensembling them showed even higher performance. These results confirm the effectiveness of general-purpose audio representation for processing heart sounds.

The contributions of this paper include 1) introducing general-purpose audio representations to the heart murmur detection task and demonstrating their effectiveness compared to SOTA, 2) showing that the representations have different performance trends and that combining them is effective in improving performance, and 3) making our code available online for future research.%\footnote{\url{https://github.com/nttcslab/m2d/tree/master/app/circor}}

\section{Experiment}
We evaluated the performance of major general-purpose audio representations and compared them to previous methods (Section \ref{sec:exp-main}),  performed ensembling to investigate the potential for further performance gain (Section \ref{sec:exp-ensemble}), and conducted ablation studies (Section \ref{sec:exp-ablation}).
For generalizability, we set up our experiments so that they follow conventional practices when solving typical audio tasks with pre-trained models, such as training with learning rate scheduling.

\subsection{Experimental Setup}
We made all experimental setups, such as tasks and evaluation metrics, based on Panah et al. \cite{Panah2023Exploring} to enable comparisons with previous studies.
We used the murmur detection task of the George B. Moody PhysioNet Challenge 2022\cite{GeorgeBMoodyPhysioNetChallenge2022} with the CirCor DigiScope heart sound dataset\cite{CirCor2022}, which is a three-class classification of cardiac murmur: Present, Absent, and Unknown (undefinable).

\vspace{0.05cm}\noindent\textbf{Audio representations.}\hspace{0.2cm}
We chose PANNs CNN14\cite{kong2020panns}, BYOL-A\cite{niizumi2021byol-a,niizumi2023byol-a}, AST\cite{gong2021ast}, and M2D\cite{niizumi2022M2D} for diversity. While these models have been commonly pre-trained on AudioSet\cite{gemmeke2017audioset}, they differ in learning methods and network architecture; PANNs pre-trained a CNN by supervised learning (SL), BYOL-A pre-trained a CNN by self-supervised learning (SSL), AST pre-trained a transformer by SL, and M2D pre-trained a transformer by SSL. Table \ref{tab:exp:models-settings}(i) shows the details of each model. We used the publicly available code and pre-trained weights of these models.

\vspace{0.05cm}\noindent\textbf{Pre-training dataset.}\hspace{0.2cm}
AudioSet\cite{gemmeke2017audioset} is a dataset of approximately 2.1M samples (approximately 5,800 h) with a 10-s duration and consists of 527 class labels. The number of samples used to pre-train each model varies because the available YouTube videos differ among research environments. For example, PANNs CNN14 used 1,934,187 samples, while M2D used 2,005,132. For pre-training, SL models (CNN14 and AST) used both audio waveforms and labels, while SSL models (BYOL-A and M2D) only used audio waveforms.

\vspace{0.05cm}\noindent\textbf{Task dataset.}\hspace{0.2cm}
We used the publicly available subset of the dataset containing samples of 963 patients from the CirCor DigiScope dataset. The data consists of samples for classes Present/Absent/Unknown of 179/695/68; it is imbalanced.
Each sample consists of multiple recordings of variable-length audio, and there are 3,163 recordings in total.
The data were split with stratification by class labels into training/validation/test sets with a proportion of 65:10:25 as in \cite{Panah2023Exploring}. We prepared three splits with different random seeds, evaluated models on these splits independently, and averaged the results as final results.

The dataset also comes with socio-demographic data, but we used only the audio data and the corresponding class labels.

\vspace{0.05cm}\noindent\textbf{Metrics.}\hspace{0.2cm}
We use weighted accuracy (W.acc)\cite{GeorgeBMoodyPhysioNetChallenge2022,Walker2022Dual} and unweighted average recall (UAR)\cite{schuller09_interspeech,Koike2020CNN14} evaluation metrics.
W.acc is a metric weighted to the classes Present and Absence:
\begin{equation}
\text{W.acc}=(5c_p+3c_u+c_a)/(5t_p+3t_u+t_a),
\end{equation}
where $c_i$ and $t_i$ for $i\in\{p\text{(resent)},u\text{(nknown)},a\text{(bsent)}\}$ are the number of correct results and true labels, respectively.
UAR is a metric that reflects performance balance among classes, and it becomes low when the recall performance of some classes is low:
\begin{equation}
\text{UAR}=1/N_c\sum\nolimits_{i\in\{p,u,a\}} r_i,
\end{equation}
where $r_i$ is recall, and $N_c$ is the number of classes, i.e., three.

\begin{figure}[tbp]
  \vspace{-5pt}
  \centering
  \includegraphics[width=0.95\columnwidth]{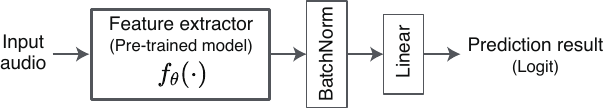}
  \vspace{-5pt}
  \caption{Network architecture. The pre-trained model $f_\theta$ is used as the feature extractor, and the output of the linear layer is the prediction result. The weights $\theta$ of the $f_\theta$ are initialized with pre-trained weight parameters, and all network layer parameters, including $\theta$, are trained in a fine-tuning fashion.}
  \label{fig:system}
  \vspace{-10pt}
\end{figure}

\vspace{0.05cm}\noindent\textbf{Network architecture.}\hspace{0.2cm}
We employed the same network architecture as in \cite{niizumi2021byol-a,niizumi2022M2D} as illustrated in Fig. \ref{fig:system}, which was used for evaluating models on the general audio tasks. The network uses a pre-trained model as a feature extractor and adds a batch normalization layer and a linear layer on top of the pre-trained model. We train all network parameters.

\vspace{0.05cm}\noindent\textbf{Training and test details.}\hspace{0.2cm}
We set the network with one of the pre-trained models as a feature extractor and trained it. After the training phase, we tested the best checkpoints on the validation set during the training to obtain the test results for the model.
We used the final evaluation result of the average of 15 experiments per model (five experiments for each of the three data splits).
All recordings were converted to a sampling rate of 16 kHz to match the model input in advance. We uniformly provided a 5-s fixed-length audio input to the model for simplicity.

For the training phase, we implemented the code based on the evaluation package {EVAR}\footnote{\url{https://github.com/nttcslab/eval-audio-repr}}.
The training/validation audio used for training was split every 5 s with a 2.5-s stride as in \cite{Panah2023Exploring}.
We used AdamW as an optimizer and trained for 50 epochs. We set the loss weights for cross-entropy loss according to sample proportions per class to deal with data imbalance. The learning rate was scheduled by cosine annealing with five epochs of warm-up, and SpecAugment\cite{specaugment} was used for data augmentation. The hyperparameters searched for each model are shown in Table \ref{tab:exp:models-settings}(ii).

For the test phase, we extended the code\footnote{\url{https://github.com/Benjamin-Walker/heart-murmur-detection}} from \cite{Walker2022Dual} to calculate UAR.
We split each recording into 5-s segments without overlap, fed them to the model to get the logits for each 5-s segment, calculated the average logit, and then obtained the per-recording results using the softmax function.
To enable direct comparison with previous studies, we employed the same heuristic label decision rule as in \cite{Panah2023Exploring,Walker2022Dual} to determine the per-patient results using the per-recording results: \textit{The prediction result is Present if any recording is classified as Present, or Unknown if any recording is classified as Unknown; otherwise, it is classified as Absent.}

\subsection{Results: Comparison with Previous Studies} \label{sec:exp-main}

\begin{table}[tb!]
%\vspace{-15pt}
\caption{Comparison with previous methods}
\label{tab:exp:result-following}
\vspace{-5pt}
\centering
\resizebox{1.0\columnwidth}{!}{%
\begin{tabular}{lrrrrr}
\toprule
       &        &        &  \multicolumn{3}{c}{Recall} \\
\cmidrule(lr){4-6}  
Model  &   W.acc &    UAR &  Present &  Unknown &  Absent \\
\midrule
\multicolumn{5}{l}{\textit{Previous studies}}  \\
Panah et al. \cite{Panah2023Exploring} & 0.80 & 0.70 & 0.86 & 0.41 & 0.83 \\
CUED\_Acoustics \cite{CUED_acoustics2022} & 0.80 & 0.68 & \textbf{0.93} & 0.34 & 0.78 \\
\midrule
\multicolumn{5}{l}{\textit{Pre-trained general-purpose audio representations}}  \\
CNN14 & 0.582 & 0.548 &      0.750 &      0.506 &     0.388 \\
BYOL-A & 0.556 & 0.556 &      0.590 &      0.573 &     0.507  \\
AST   & 0.654 & 0.672 &      0.744 &  \textbf{0.769} &     0.505 \\
M2D & \textbf{0.832} & \textbf{0.713} &      0.911 &      0.361 & \textbf{0.868} \\
\bottomrule
\end{tabular}
}
\vspace{-10pt}
\end{table}

Table \ref{tab:exp:result-following} compares the results with previous studies and shows that M2D, the recent model, outperforms previous results.
M2D pre-trains a representation by SSL using a masked prediction task, an effective task for learning useful representations, which is especially popular in the natural language and image domains. The results show that the representation is also effective in heart murmur detection.

Notably, the experiments achieved high performance by simply using common data augmentation and learning techniques and solving the task in the same way as general audio tasks. Prior work \cite{Panah2023Exploring} used wav2vec 2.0\cite{baevski2020wav2vec2} to pre-train on the CirCor DigiScope dataset by masked prediction-based SSL, and \cite{CUED_acoustics2022} adapted a hidden semi-Markov model for heart murmur detection. In contrast, as shown by the results in Table \ref{tab:exp:result-following}, M2D outperforms the previous studies, even without requiring the domain techniques used in those studies.

We observe several performance trends when focusing on the class-specific recall performance of each model. In the Unknown class, the performance of M2D is inferior compared to other models. Conversely, M2D exhibits superior performance in the Absent class, while other models show lower performance. The possible explanation for these differences could be the data augmentation, which M2D does not use, whereas the others do.
While models other than M2D underperform the W.acc and UAR results compared to the previous methods, their performance trends suggest that an ensemble of M2D and other models could enhance overall performance.

CNN pre-trained models (CNN14 and BYOL-A) perform poorly compared to previous studies, while transformer-based models (AST and M2D) show high performance. The previous study\cite{Panah2023Exploring} also achieved high performance by applying the transformer-based speech model wav2vec 2.0\cite{baevski2020wav2vec2}, suggesting that a transformer is advantageous for the task.
Since a CNN is generally good at handling local features, while a transformer is good at handling global relationships of the entire input sequence\cite{SurveyViT}, the task may require better handling of global relationships.

Comparing AST and M2D, which share a very close transformer architecture, M2D is an SSL that does not rely on labels, while AST learns AudioSet's labels. The results suggest that the task differs from the classification task of AudioSet; thus, AST representation is not effective enough, and the M2D representation is effective thanks to the pre-training by a task-agnostic SSL.

\subsection{Results: Ensemble of Models} \label{sec:exp-ensemble}

\begin{table}[tb!]
\vspace{-5pt}
\caption{Ensemble results of two models}
\label{tab:exp:result-ensemble}
\vspace{-5pt}
\centering
\resizebox{1.0\columnwidth}{!}{%
\begin{tabular}{lrrrrr}
\toprule
       &        &        &  \multicolumn{3}{c}{Recall} \\
\cmidrule(lr){4-6}  
Ensemble  &   W.acc &    UAR &  Present &  Unknown &  Absent \\
\midrule
CNN14+M2D     & 0.829 & \underline{0.719} &      0.898 &  \underline{0.391} & \underline{\textbf{0.868}} \\
BYOL-A+M2D    & 0.817 & \underline{0.721} &      0.870 &  \underline{0.432} & 0.862 \\
AST+M2D       & \underline{\textbf{0.832}} & \underline{\textbf{0.733}} & \textbf{0.899} & \underline{\textbf{0.438}} &     0.862 \\
\bottomrule
\addlinespace[0.05cm]
\multicolumn{5}{l}{\scriptsize{Underlined results are equal to or better than M2D results in Table \ref{tab:exp:result-following}.}} \\
\end{tabular}
}
\vspace{-10pt}
\end{table}

We conducted experiments to explore the possibility of improving performance by ensembling the predictions of multiple models.
For simplicity, we experimented with ensembles of two models, with M2D fixed as one of them.
We used the predictions of each model for each recording obtained from the experiments in Section \ref{sec:exp-main}.
Then, we obtained the prediction results using the average class probability of the predictions of each model we ensembled.
To obtain the ensembled prediction for each recording, we averaged a total of 25 predictions, which are all combinations of five predictions for each of the two models from the ensemble.

The results in Table \ref{tab:exp:result-ensemble} show that the AST+M2D ensemble can improve the detection of Unkown and, consequently, the overall metrics W.acc and UAR.
The CNN14+M2D and BYOL-A+M2D results show that they can improve the Unkown detection performance, although the overall performance is slightly degraded.
These results indicate that higher performance can be achieved by combining models, such as M2D, which has better overall performance, and other models, which have strong Unkown performance.

\subsection{Ablation Study} \label{sec:exp-ablation}

We conducted experiments as an ablation study without using the label decision rule, without the data augmentation, and without pre-training M2D to investigate the contribution of each to the final performance.

Without using the label decision rule, we calculated the average prediction probabilities for multiple recordings of a given patient and then used them as the prediction result for that patient.
In the results shown in Table \ref{tab:exp:result-not-following}(a), the performance of Present dropped while that of Absent improved for all models; these results reflected the non-use of the label decision rule where the prediction for a patient is Present if there is even one recording with the highest probability of Present. However, the performance of W.acc and UAR became higher for all models except M2D, indicating that these models perform well as they are without the heuristic rule, and the final patient's results should be averaged directly from the model's predictions.

Notably, M2D, the only one achieving SOTA performance among the models, decreased the overall performance without the rule, indicating that M2D had a class recall performance balance similar to the previous methods.

\begin{table}[tb!]
%\vspace{-15pt}
\caption{Ablation study results}
\label{tab:exp:result-not-following}
\vspace{-5pt}
\centering
\resizebox{1.0\columnwidth}{!}{%
\begin{tabular}{lrrrrr}
\toprule
       &        &        &  \multicolumn{3}{c}{Recall} \\
\cmidrule(lr){4-6}  
Model  &   W.acc &    UAR &  Present &  Unknown &  Absent \\
\midrule
\multicolumn{6}{l}{\textit{(a) Without using the label decision rule \cite{Panah2023Exploring}}}  \\
CNN14  & 0.611$\uparrow$ & 0.604$\uparrow$ & 0.544$\downarrow$ &   0.553$\uparrow$ &   0.715$\uparrow$ \\
BYOL-A & 0.569$\uparrow$ & 0.598$\uparrow$ & 0.409$\downarrow$ &   0.627$\uparrow$ &   0.759$\uparrow$ \\
AST  & 0.673$\uparrow$ & 0.705$\uparrow$ &   0.579$\downarrow$ &   0.769\hspace{0.15cm} &  0.766$\uparrow$ \\
M2D & 0.796$\downarrow$ & 0.684$\downarrow$ & 0.794$\downarrow$ &  0.318$\downarrow$ &     0.940$\uparrow$ \\
 & \multicolumn{5}{c}{\scriptsize{$\uparrow/\downarrow$ show the difference from Table \ref{tab:exp:result-following}}}  \\
\midrule
\multicolumn{6}{l}{\textit{(b) Without using the data augmentation (SpecAugment)}}  \\
CNN14 (no aug.) & 0.374 & 0.374 &      0.550 &      0.447 &     0.126 \\
BYOL-A (no aug.) & 0.536 & 0.524 &      0.630 &      0.522 &     0.420 \\
AST (no aug.) & 0.670 & 0.617 &      0.772 &      0.490 &     0.590 \\
\midrule
\multicolumn{6}{l}{\textit{(c) Using random initial weights (no pre-training)}}  \\
M2D (random) & 0.595 & 0.536 &      0.547 &      0.325 &     0.737 \\
\bottomrule
\end{tabular}
}
\vspace{-10pt}
\end{table}

As shown in Table \ref{tab:exp:result-not-following}(b), the overall performance deteriorated without data augmentation (SpecAugment), indicating that data augmentation plays a significant role in improving performance in this case of the small number of data in the task. The exception is M2D, with which preliminary experiments have shown that SpecAugment cannot improve performance.
This may also indicate that M2D had pre-trained a representation directly applicable to the task.
Overall, the need for data augmentation also depends on the model used.

The case without pre-training was experimented with M2D, and the results are shown in Table \ref{tab:exp:result-not-following}(c). In this case, the transformer weights of M2D were randomly initialized and trained from scratch. %Note that all experiment setups in this study are fine-tuning that train all model parameters.
The results shown in the table are very poor compared to those in Table \ref{tab:exp:result-following}, confirming that pre-training useful representations is essential.

\section{Conclusion}
This study investigated the effectiveness of general-purpose audio representation in the heart murmur detection problem. Among the representations pre-trained on a large general audio dataset, the recent self-supervised learning model M2D outperformed the previous studies, while the other models showed remarkable performance in detecting one of the classes, Unknown, for which M2D performance suffered. The ensemble of these models showed even better performance, indicating further potential for performance contribution.
Our results demonstrate that the pre-trained general-purpose audio models are effective even without domain-specific adaptation.
Furthermore, these results also show that the pre-training on a large dataset is also vital in this domain, the same as in the other domains, such as speech, opening the door for further advancements. Our code is available online\footnote{\url{https://github.com/nttcslab/m2d/tree/master/app/circor}}

\bibliographystyle{IEEEtran}
\bibliography{refs}

\end{document}